\documentclass[aps,showpacs,
amssymb,endfloats]{revtex4}
\usepackage{bm,amsfonts}
\begin{document}

\title {Electromagnetically induced transparency for $\Lambda$ - like systems
with a structured continuum}
\author{A. Raczy\'nski}\email{raczyn@phys.uni.torun.pl}\author{M. Rzepecka}\author{J. Zaremba}

\affiliation{Instytut Fizyki, Uniwersytet Miko\l aja Kopernika,
ulica Grudzi\c{a}dzka 5, 87-100 Toru\'n, Poland,}

\author{S. Zieli\'nska-Kaniasty}
\affiliation{Instytut Matematyki i Fizyki, Akademia
Techniczno-Rolnicza, Aleja Prof. S. Kaliskiego 7, 85-796
Bydgoszcz, Poland.}

\begin{abstract}
Electric susceptibility of a laser-dressed atomic medium is
calculated for a model $\Lambda$-like system including two lower
states and a continuum structured by a presence of an autoionizing
state or a continuum with a laser-induced structure. Depending on
the strength of a control field it is possible to obtain a
significant reduction of the light velocity in a narrow frequency
window in the conditions of a small absorption. A smooth
transition is shown between the case of a flat continuum and that
of a discrete state serving as the upper state of a $\Lambda$
system.
\\
\pacs{42.50.Gy, 32.80.-t}
\end{abstract}
\maketitle
\newpage
It has been known for a long time that atoms in the $\Lambda$
configuration irradiated by two laser fields may exhibit peculiar
dynamical properties. In particular the population may be trapped
in the so-called dark state, being a combination of the two lower
states. It constitutes a basis for subtle coherent dynamical
manipulations like an efficient population transfer (stimulated
Raman adiabatic passage - STIRAP), robust against any broadening
of the upper state \cite{b1,b2}. Effects of such a type are
present also in the case in which the upper state of a $\Lambda$
configuration has been replaced by a continuum \cite {b3,b4,b5}
or, in strong laser fields, by a set of coupled continua
\cite{b6}.

Dynamical effects in particular atoms are reflected in unusual
light propagation effects in atomic media. One of the most
important propagation effects in a medium of atoms in the
$\Lambda$ configuration is the electromagnetically induced
transparency (EIT), which consists in making the medium
transparent for a weak probe laser beam by irradiating it by a
strong copropagating control beam \cite{b7,b8}. Instead of an
absorption line there appears a transparency window and a normal
dispersion. By switching the control field off when the pulse is
inside the sample one can slow the light pulse down and finally
stop or store it \cite{b9,b10,b11}. The photon energy of the
signal is transferred to the control beam while the information
about the pulse is written down in the form of a coherence
(nondiagonal element of the density matrix) between the two lower
states. Switching the signal on again results in reading the
information out: the pulse is reconstructed in a coherent way.

A natural question arises whether similar propagation effects may
occur in the case of a continuum serving as an upper state. Some
answer has been given by van Enk {\em et al.} \cite{b12,b13}, who
considered a general case of two beams propagating inside the
medium. Their conclusion was that a necessary condition for EIT
was that the asymmetry parameter in the continuum model was zero.
In their case the losses due to photoionization are of second
order with respect to the probe field. Thus, in the linear
approximation with respect to the probe beam and neglecting
propagation effects for the strong control beam, one still may
apply the standard approach in which the probe propagation is
discussed in terms of an atomic susceptibility. Below we give
analytic expressions for the susceptibility in the case of the
bound-continuum dipole matrix elements being modeled according to
Fano autoionization theory \cite{b14}. By changing the Fano
asymmetry parameter we pass from the case of a flat continuum to
that of a discrete and broadened bound state. We examine the shape
of the transparency window depending on the amplitude of the
coupling field.

We consider a generalization of a $\Lambda$ system in which the
upper state is replaced by a continuum: an atomic system including
two lower discrete states $b$ (initial) and $c$ and a continuum
$E$ coupled with the state $b$ by a weak probe field
$(\frac{1}{2}\epsilon_{1}(z,t) \exp[i(k_{1}z-\omega_{1}t)]+c.c$
and with the state $c$ with a relatively strong control field
$\epsilon_{2} \cos\omega_{2}t$. The continuum may have a Fano
density of states, due to a configurational coupling with an
autoionizing state or to a structure induced by another laser
\cite{b15}. As usually in EIT, the control field, the propagation
effects for which are neglected, dresses the atomic medium to
create new conditions for the propagation of the probe pulse. The
evolution of the atomic system is described by the von Neumann
equation, which after transforming-off the rapidly oscillating
terms and making the rotating-wave approximation is reduced in the
first order perturbation with respect to the probe field to the
set of the following equations for the density matrix $\sigma$
\begin{eqnarray}
i\hbar\dot{\sigma}_{Eb}=(E-E_{b} -\hbar\omega_{1})\sigma_{Eb}-
\frac{1}{2}d_{Eb}\epsilon_{1}-\frac{1}{2}d_{Ec}\epsilon_{2}\sigma_{cb},\nonumber\\
i\hbar\dot{\sigma}_{cb}=(E_{c}+\hbar\omega_{2}-E_{b}-\hbar\omega_{1}
-i\gamma_{cb})\sigma_{cb} -\frac{1}{2}\epsilon_{2}^{*}\int
d_{cE}\sigma_{Eb} dE.
\end{eqnarray}
In the above equation $d$ is the dipole moment and $\gamma_{cb}$
is a phenomenological relaxation rate for the coherence
$\sigma_{cb}$. Stationary solutions of Eqs (1) can be found by
first expressing $\sigma_{cb}$ in terms of $\sigma_{Eb}$ and then
by solving the integral equation for the latter. The component of
the polarization of the medium connected with the $b-E$ coupling
is
\begin{equation}
P^{+}(\omega_{1})=N\int d_{bE}\sigma_{Eb}dE =
\epsilon_{0}\chi(\omega_{1})\epsilon_{1}(\omega_{1}),
\end{equation}
with $\epsilon_{0}$ being the vacuum electric permittivity, $N$ -
the atom density and with the medium susceptibility given by
\begin{equation}
\chi(\omega_{1})=-\frac{N}{2\epsilon_{0}} \left(R_{bb}+\frac{1}{4}
\frac{\epsilon_{2}^{2}R_{bc}R_{cb}}
{E_{b}+\hbar\omega_{1}-E_{c}-\hbar\omega_{2}+i\gamma_{cb}
-\frac{1}{4}\epsilon_{2}^{2}R_{cc}} \right).
\end{equation}
The function $R_{ij}(\omega_{1})$, $i,j=b,c$, is given by
\begin{equation}
R_{ij}(\omega_{1})=\int_{0}^{\infty} \frac{d_{iE}d_{Ej}
}{E_{b}+\hbar\omega_{1}-E+i\eta}dE,
\end{equation}
where $\eta\rightarrow 0^{+}$ assures the correct behavior at the
branch cut. The presence of an autoionizing state coupled with the
continuum results in a modification of the density of the
continuum states, which leads to the following Fano formula for
the bound-structured continuum dipole matrix element
\begin{equation}
|d_{iE}|^{2}=B_{i}^{2}\frac{(E-E_{a}+q\gamma)^{2}}{(E-E_{a})^{2}+\gamma^{2}},
\end{equation}
where $E_{a}$ and $\gamma$ characterize the position and width of
the autoionizing resonance, $q$ is the Fano asymmetry parameter
and $B_{i}$ is the bound-flat continuum matrix element. We assume
that the matrix elements for the states $b$ and $c$ differ only by
a constant factor $B_{b,c}$, i.e. the presence of the autoionizing
state is the only reason for the continuum to exhibit any
structure. For real systems the energy dependence of the matrix
element is more complicated, in particular an additional form
factor is necessary to account for a threshold behavior or a
correct asymptotic properties. Taking such a form factor into
account is not essential in an approach like ours, in which we
concentrate on the frequency dependence of the susceptibility in a
narrow band close to the resonance.

If the lasers are tuned sufficiently far from the ionization
threshold and so is the localization of the autoionizing state we
can extend the lower limit of the energy integral to $-\infty$ and
get from Eqs (4) and (5)
\begin{equation}
R_{ij}(\omega_{1})=B_{i}B_{j}\pi \frac
{\gamma(\hbar\omega_{1}-E_{a})(q^{2}-1)+2q\gamma^{2}-i(\hbar\omega_{1}-E_{a}+q\gamma)^{2}}
{(\hbar\omega_{1}-E_{a})^2+\gamma^{2}}
\end{equation}

It is possible to evaluate the width of the structure in the
susceptibility given by Eqs (3) and (6) if it is significantly
smaller than an autoionizing width, i.e $\gamma>>q\omega_{1}$. The
width for $q>>1$ is in atomic units of order of
$\epsilon_{2}^{2}q^{2}$. This result is  to be compared with the
width of the transparency window for EIT in a typical $\Lambda$
system, i.e. one with a discrete broadened upper state. The width
of the latter window is in atomic units of order of
$\epsilon_{2}^{2}/\Gamma$, $\Gamma$ being the relaxation rate for
the nondiagonal element of the atomic coherence. One can see that
for the same control fields the transparency window here is much
more narrow. For a typical value of $\Gamma\approx 10^{-8}$ a.u.
we need $q\approx 10^{4}$ to make the windows in the two cases
comparable. The light absorption in the case of the continuum is
by a few orders of magnitude weaker (again except for very large
$q$) than in the case of a discrete upper level but due to the
transparency window being narrow the dispersion can be very steep,
so a significant light slowdown is possible.

As an illustration of the above formalism we show the atomic
susceptibility calculated from Eqs (3) and (6). We have taken
$\gamma=10^{-9}$ a.u.$\approx 6.6$ MHz. The values of the field
amplitude $\epsilon_{2}$ ranged from $10^{-9}$ to $10^{-6}$ a.u.,
i.e. the width $2\pi \epsilon_{2}^{2}B_{j}^{2}/4$ of the lower
states, induced by the bound-continuum coupling for the flat
continuum (without the Fano factor) would be of order of
$10^{-8}\gamma$ to $10^{-2}\gamma$. The values of the parameters
$B_{j}$ were taken arbitrarily $B_{1}=2$ a.u., $B_{2}=3$ a.u. The
atomic density was $N=0.67\times10^{12}$ cm$^{-3}$. The asymmetry
parameter $q$ was of order of 10-100. The position of the states
was taken $E_{a}=-E_{b}/2=-E_{c}=0.1$ a.u.= 2.72 eV. The
relaxation rate $\gamma_{cb}$, possibly due to the spontaneous
emission from the level $c$ or to an incoherent ionization channel
by the field $\epsilon_{2}$ from the level $b$, was neglected.

In Fig. \ref{fig1} we show the medium susceptibility for $q$=10
and $\epsilon_{2}=0$. As expected, a normal dispersion is observed
except close to the resonance, together with a single absorption
line. Note that the maximum values of both the real and imaginary
parts of $\chi$ are smaller than $10^{-9}$. Fig. \ref{fig2} shows
the susceptibility after switching on a relatively strong control
field $\epsilon_{2}=10^{-6}$ a.u. Note the transparency window of
the width of order of $\gamma$ and the corresponding interval of a
normal dispersion at a resonance frequency. Due to the Fano
asymmetry the curves in the figure lack symmetry too. The group
index $n_{g}=1+\frac{\omega_{1}}{2}\frac{d}{d\omega_{1}} {\rm
Re}\chi(\omega_{1})$, being the factor by which the probe signal
velocity is reduced, is only about 1.1. Figs \ref{fig3} and
\ref{fig4} show how the frequency window is narrowed for smaller
control fields. The widths are of order of $10^{-2}\gamma$ and
$10^{-4}\gamma$ respectively, while the corresponding group
indices are of order of 10 and 1000. The spectral width of the
laser beam needed in the latter case is however not yet
accessible.  The effects of a reduced absorption and a steep
normal dispersion are made stronger if the Fano parameter $q$ is
increased. In Fig. \ref{fig5} we show the suspecptibility for
$q=100$ and $\epsilon_{2}=10^{-9}$ a.u. ($q$ has been increased
and $\epsilon_{2}$ has been decreased by the factor of 10 compared
with Fig. \ref{fig4}). Note that absolute values of $\chi$ are
increased by two orders of magnitude compared with those of Fig.
\ref{fig4} while the transparency window is the same. This means
that the group index is about $10^{5}$.

We have investigated the optical properties of an atomic medium in
the $\Lambda$ configuration in which the upper state is replaced
by a continuum. Such a system exhibits properties similar to those
of a typical $\Lambda$ system, except that the orders of their
magnitude are changed. Here the windows of a reduced absorption
(which is in general much smaller for the systems examined here)
and of a normal dispersion are much more narrow, if the control
fields are not made stronger. The dispersion curve in the
transparency window may still be steep, which gives rise to a
significant light slowdown. By increasing the Fano asymmetry
parameter by a few orders of magnitude we smoothly pass to the
case of a typical $\Lambda$ system.

\begin{acknowledgments}
The work has been supported in part by the Committee for
Scientific Research, grant No. 1 P03B 010 28. The subject belongs
to the scientific program of the National Laboratory of AMO
Physics in Toru\'n, Poland.
\end{acknowledgments}

\newpage

\begin{figure}
\caption {\label{fig1} The real and imaginary parts of the
susceptibility in the case of $\epsilon_{2}=0$ and $q=10$. Zero
frequency corresponds to the resonance with the autoionizing
state. }
\end{figure}

\begin{figure}
\caption {\label{fig2} As in Fig.\ref{fig1} but for
$\epsilon_{2}=10^{-6}$ a.u.}
\end{figure}

\begin{figure}
\caption {\label{fig3} As in Fig.\ref{fig1} but for
$\epsilon_{2}=10^{-7}$ a.u.}
\end{figure}

\begin{figure}
\caption {\label{fig4} As in Fig.\ref{fig1} but for
$\epsilon_{2}=10^{-8}$ a.u.}
\end{figure}

\begin{figure}
\caption {\label{fig5} As in Fig.\ref{fig1} but for
$\epsilon_{2}=10^{-9}$ a.u. and $q=100$.}
\end{figure}
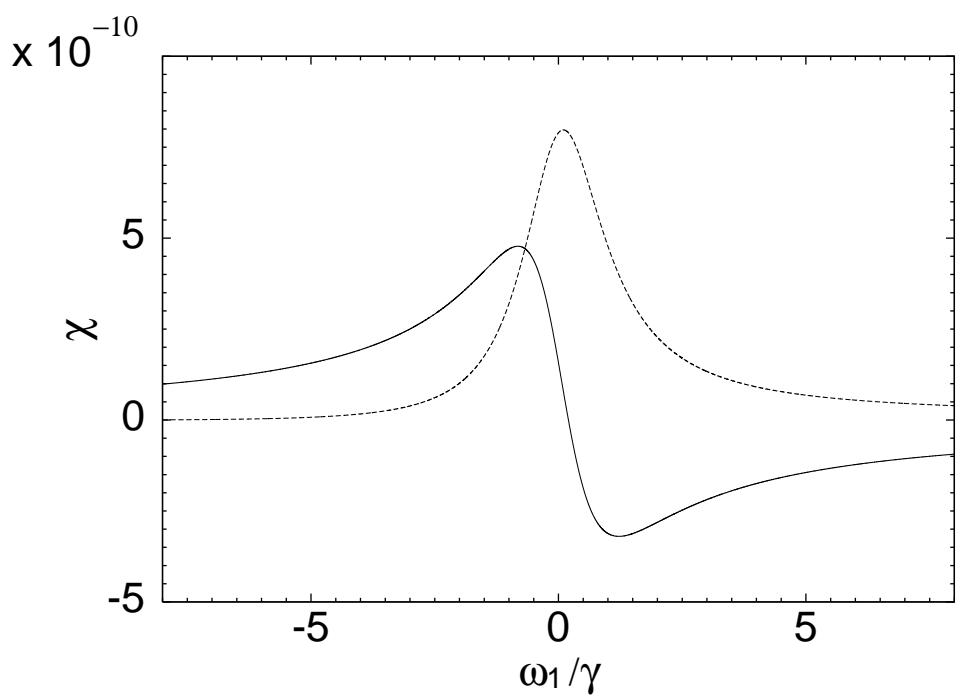
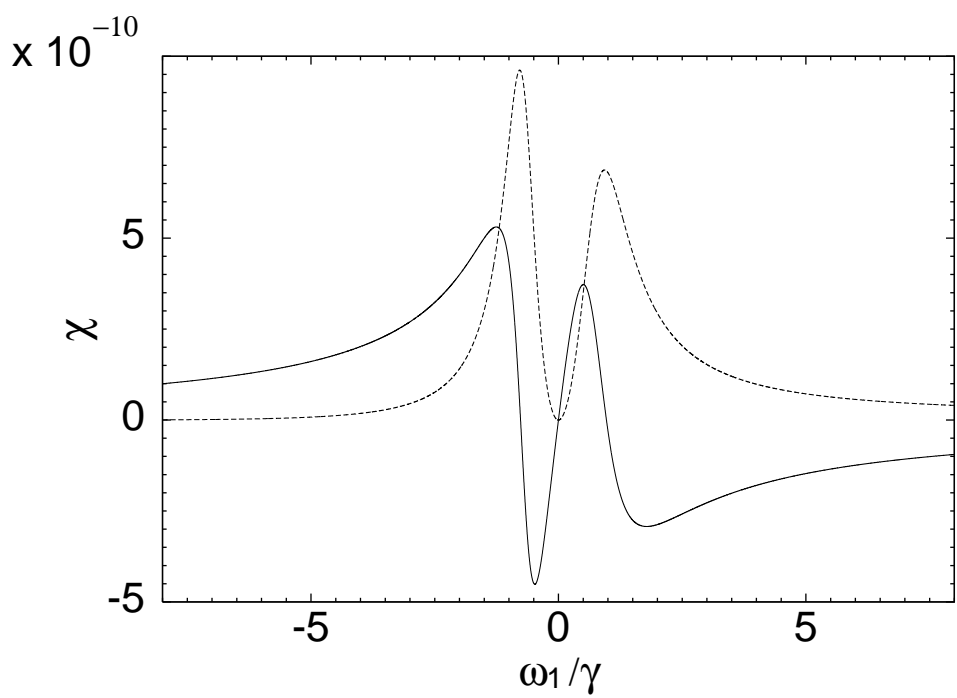
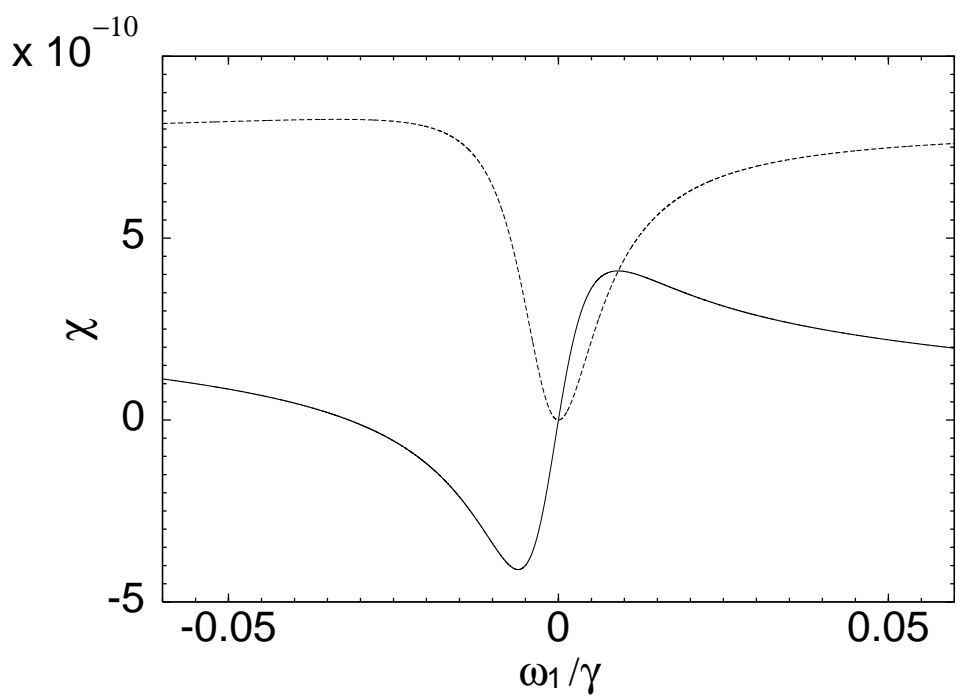
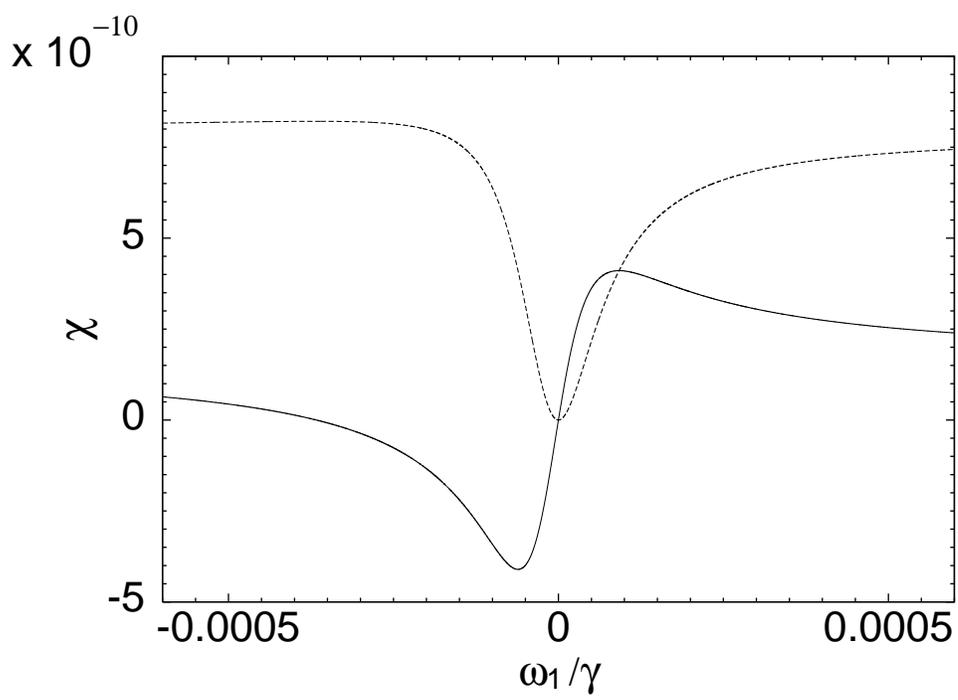
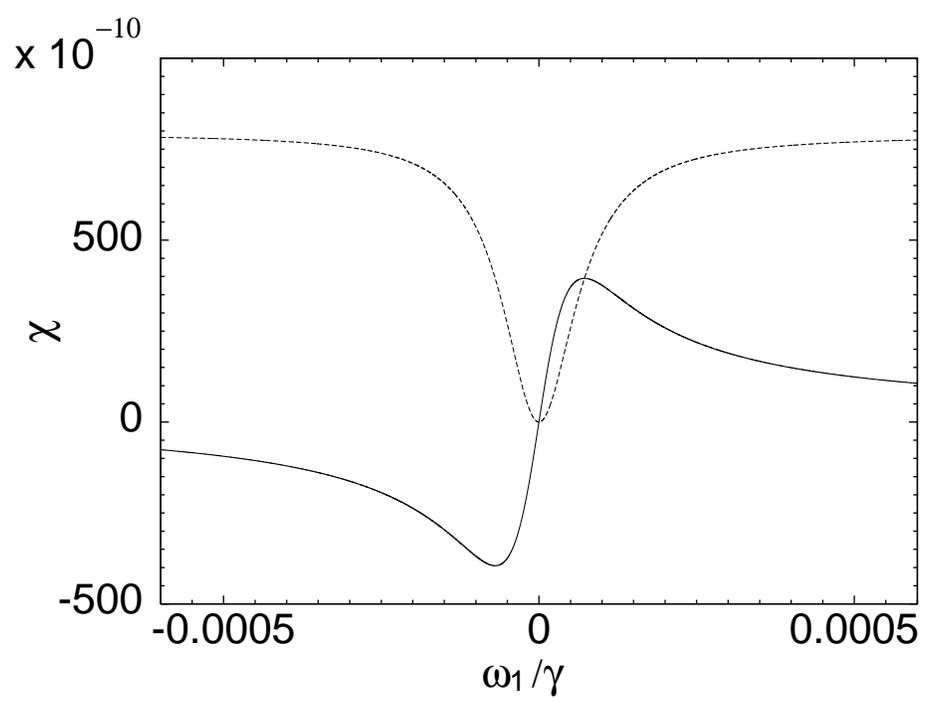
\end{document}